\documentstyle[psfig,prb,aps]{revtex}
\font\ee=msbm8 scaled \magstep1
\font\ee=msbm8 scaled \magstep1

\tightenlines
\date{\today}
\begin{document}
\vskip.5in minus.2in
\begin{center}
{\bf \Large Nonlocal, noncommutative picture in quantum mechanics  \\ 
and distinguished canonical maps}
\vskip.5in minus.2in

{T. Hakio\u{g}lu}
\\[5mm]
\begin{center}
{\em Physics Department, \\
Bilkent University, \\ 
06533 Ankara, Turkey 
}
\end{center}
\end{center}
\vskip2truecm
\begin{abstract}
Classical nonlinear canonical (Poisson) maps have a distinguished role in 
quantum mechanics. They act unitarily on the quantum  
phase space and generate $\hbar$-independent quantum 
canonical maps. It is shown that such maps act in the noncommutative 
phase space under the classical covariance. 
A crucial observation made is that under the classical covariance 
the local quantum mechanical picture can become 
nonlocal in the Hilbert space. This nonlocal picture is made equivalent    
by the Weyl map to a full noncommutative picture  
in the phase space formulation of the theory. The connection between the 
entanglement and nonlocality of the representation is explored and specific 
examples of the generation of entanglement by using the concept of generalized 
Bell states are provided.  
That the results have direct application in generating vacuum soliton 
configurations in the recently popular scalar 
field theories of noncommutative coordinates is also demonstrated.   
\end{abstract}

~~~~~~~~~~~~~~~~03.65.Sq (Semiclass. theories and appl's.),  

~~~~~~~~~~~~~~~~03.65.Fd (Algebraic methods)
\twocolumn
\vskip0.5truecm
\subsection{Introduction}
Recently increased activity in noncommutative (field and gauge)  
theories\cite{nc} and the natural observation of
noncommutative spatial coordinates in string physics\cite{st} arose a flurry
of interest in a certain nonstandard extension of the quantum mechanics 
within this noncommutative picture.\cite{ncqm}
In these theories the fundamental entity, the field, acquires    
an operator character by its dependence
on the noncommutative coordinates (NCC). The NCC can arise in
certain physical limits of a quantum theory. Recently,  
Bigatti and Susskind\cite{BS} have suggested the quantum model in two  
space dimensions of a charged particle in the plane interacting with       
a perpendicular magnetic field in the limit the field strength goes to 
infinity to observe noncommutativity in the plane coordinates.  
Considering that the noncommutative gauge theories are    
related to the ordinary ones by certain   
transformations\cite{st} it is reasonable to inquire whether more general 
results can be obtained on the nature of the interrelationship and the 
possibility of invertible transformations 
between the noncommutative and the ordinary formalisms. 
For instance, in the case of gauge theories the {\it ordinary} and the 
noncommutative theories are related by a {\it gauge 
equivalence preserving} map reminiscent of a contact 
transformation between the fields and the gauge parameters 
vice versa.\cite{st} 
On the other hand, the {\it phase space} is spanned by the  
generalized coordinates and the formulation of quantum mechanics  
in the phase space explicitly has features of noncommutativity. 
This can be used as  
a natural testing ground to understand whether more general transformations 
can be found between the ordinary and noncommutative pictures. 
In our context here    
the noncommutative generalized coordinates   
$\hat{z}=(\hat{z}_1,\hat{z}_2)$ are the {\it generalized
momentum} and {\it position} operators respectively satisfying 
$[\hat{z}_j,\hat{z}_k]=i\theta_{j\,k}$ where $j,k=(1,2)$ and the  
noncommutativity parameter is $\theta_{j\,k}=\hbar\,J_{j\,k}$ with 
$J_{j\,k}$ describing the symplectic matrix $\pmatrix{0 & 1 \cr -1 & 0\cr}$. 

The equivalence of the one dimensional phase space to the 
two dimensional noncommutative geometry approach is already known. The 
generalization of the ideas presented here to noncommutative geometries 
may therefore be direct. For instance, in the case of Landau model this 
equivalence was shown 
in Ref.\,[5]. More recently it was proposed that there is an interesting 
connection between the quantum Hall effect and the noncommutative matrix 
models.\cite{suskind2,poly1} 
The noncommutative picture can be invertibly converted into a bilocal 
coordinate picture by the Weyl map. Once a bilocal representation is obtained  
the (non)separability of these representations becomes a crucial property. 
When the BLC dependence is separable  
the corresponding noncommutative picture reduces to the standard 
(Schr\"{o}dinger) picture. On the other hand,    
the nonseparability of the dependence on the bilocal coordinates requires  
a full phase space formalism (which may not be reducible to 
a pure coordinate or momentum 
representation as in the standard Schr\"{o}dinger picture).  
This full treatment of the noncommutative phase space and the corresponding 
nonlocal function space suggest an extended frame 
above the standard formulation of quantum mechanics. 
The purpose of this letter is to explore this extended quantum mechanical 
picture. 

In section B the basic connections between the BLC and the phase space 
representations are established by the Weyl map and the question of the 
separability versus nonseparability in the BLC picture is studied. 
It is shown that the separable bilocal solutions are actually local and 
they can be fitted in the standard Schr\"{o}dinger picture.  
The nonlocal solutions arise from the nonseparable bilocal coordinate  
dependence and they fit in the extended picture. 
In section C, it is shown that 
under the action of the standard linear Hilbert space operators the 
separable and the 
nonseparable functions are disjoint. In section D we examine the phase 
space approach in this extended quantum mechanical picture. By reaching beyond 
the standard Hilbert space operator methods, we find distinguished  
unitary (in the noncommutative -phase- space) isomorphisms (canonical maps) 
joining the separable and the nonseparable sectors. In the view of  
the continuous canonical transformations, these isomorphisms 
appear to be induced by the classical canonical (Poisson) 
maps. In section E a connection between the nonlocality and the 
entanglement is examined in the concept of {\it generalized Bell states} is 
introduced.  
Specific nonlinear Poisson maps are given in section E that produce these
states from the pure states.
Recently the generalized Bell states have been observed 
in the vacuum soliton solutions of the noncommutative field theories 
and this concept may be required in building a genuin equivalence scheme for 
the set of all noncommutative soliton solutions. This particular application 
is timely and relevant arising as an effect of the Poisson maps which is also  
discussed in a paragraph therein. The section F  is a schematical   
illustration and summary of the results in the work. 

\subsection{The extended quantum mechanical picture} 
The ordinary quantum
mechanics is the sector of the field theory supporting a single
degree of freedom. We adopt an operator-like approach in this   
description\cite{poly,chai}    
that will be referred to as the {\it waveoperator}.  
The waveoperator is a complex functional of the noncommutative generalized 
coordinates which will be denoted by   
$\hat{\rho}=\rho(\hat{z})$. 
In this representation quantum mechanics
is formulated by extremizing the action\cite{poly}
\begin{eqnarray}
{\cal S}&=&\int\,dt\,Tr\Bigl\{
i\hbar\,\hat{\rho}^{\dagger}\,\partial_t\,\hat{\rho}-
\hat{\rho}^{\dagger}\,\hat{\cal H}\,\hat{\rho}\Bigr\}~ \nonumber \\
\hat{\cal H}&=&{\cal H}(\hat{z}_1,\hat{z}_2)=
{\hat{z}^2_1 \over 2m}+V(\hat{z}_2)
\label{action.1}
\end{eqnarray}
with respect to some coordinate space description ${\rho}(y,x)=\langle y \vert 
\hat{\rho}\vert x\rangle$. Here $\hat{\cal H}$ is a Hamiltonian operator 
in the Hilbert space and  
$Tr$ stands for the trace in the position  
basis $\vert x\rangle$ (i.e. $\hat{z}_2\,\vert x\rangle=x\vert x\rangle$). 
Throughout the paper, the waveoperator is  
time dependent although we will not write it explicitly.


Evaluating the action by considering the trace
in this position basis, and minimizing it   
with respect to ${\rho}^*(x,y)=
\langle y\vert\hat{\rho}\vert x\rangle$ we obtain  
\begin{equation}
i\hbar\,\partial_t\,{\rho}(y,x)=
\int\,du\,h(y,u)\,\rho(u,x)
\label{action.3}
\end{equation}
where $\rho(y,x)$ is the representation of the waveoperator in the 
{\it bilocal coordinates} (BLC) $y,x$. Equivalently the bilocal $\rho(y,x)$ 
can be mapped isomorphically onto a doubled Hilbert space ${H\times H}$ and 
this procedure is known as the Gel'fand-Naimark-Segal construction. The 
$H \times H$ representations have been noticed recently to be of use in the 
phase space representations of the quantum canonical Lie algebras\cite{EG}. 
In this 
scheme $\rho(y,x)$ can be referred as the wavefunction in this doubled Hilbert 
space\cite{foot1}. Similarly, $h(y,u)=\langle y\vert {\cal H} \vert u\rangle$   
is the Hamiltonian in BLC. Eq.\,(\ref{action.3}) will be referred as the 
bilocal Schr\"{o}dinger equation.    

Note that, (\ref{action.3}) does not have unique solutions. The nonuniqueness 
arises from the indeterminacy in the $x$ dependence. In order to show this 
indeterminacy in (\ref{action.3}) we assume that one solution is separable,   
e.g. $\rho(y,x)=\psi_1(y)\,\psi_2(x)$ where both $\psi_k~,~~k=1,2$ 
have finite norm\cite{remark1}. Here $\psi_2(x)$ must be time independent. 
Inserting this in (\ref{action.3}) one  
obtains the standard Schr\"{o}dinger equation for $\psi_1$ whereas since  
$\psi_2$ is time independent, it appears as a multiplicative function on 
both sides and thus it cannot be determined by the equation. 
In order to reach beyond this standard  
representation the nonseparable solutions of (\ref{action.3})   
in terms of the BLC, e.g. $\rho(y,x) \ne \psi_1(y)\,\psi_2(x)$ are of 
interest. 

The separable and nonseparable sectors in the space of bilocal functions 
can be connected to each other
by certain types of canonical maps. This can be best understood 
if we use the Weyl map\cite{Weyl} 
to transform $\rho(y,x)$ into a function 
$\tilde{\rho}(z)$ where $z=(z_1,z_2)$ 
are the generalized coordinates in some noncommutative space 
${\cal Z}_{\star}$. Note 
that throughout the paper a generic dependence on $(z_1,z_2)$ will be denoted 
by $z$ and functions of $z$ will be hatted with a tilde.
Also particularly in this work, $z_1,z_2$ describe the phase space 
coordinates, e.g. the generalized canonical   
coordinate and momentum respectively. The standard approach to the phase space 
quantum mechanics is to use the Weyl correspondence which is an analytic and 
invertible map from an arbitrary Hilbert space 
operator $\hat{O}$ to a function $\tilde{o}(z)$ in ${\cal Z}_{\star}$. 
This correspondence will be denoted by 
${\cal W}: \hat{O} ~\leftrightarrow ~ \tilde{o}(z)$. 
Since a Hilbert space operator can also be represented by a bilocal function  
as $\hat{O} ~\leftrightarrow ~ \langle y\vert\hat{O}\vert x\rangle=o(y,x)$, 
the combination of these two 
is also a well-defined (Weyl) map $W: o(y,x) ~\leftrightarrow ~ \tilde{o}(z)$.  
The Weyl map is explicitly given by
\begin{equation}
o(y,x)=\int\,{d^2 z \over 2\pi\hbar}\,{\cal K}_z(y,x)\,
\tilde{o}(z)~.
\label{action.5}
\end{equation}
Here ${\cal K}_z(y,x)=e^{iz_1(x-y)/\hbar}\,\delta(z_2-{x+y \over 2})$ is  
the invertible Weyl kernel. The function $\tilde{o}(z)$ is the standard Weyl
symbol of $\hat{O}$ in the phase space ${\cal Z}_{\star}$. In this 
context, $o$ and $\tilde{o}$ are two different representations of 
the Hilbert space operator $\hat{O}$. We therefore have the   
following triangle diagram  
\begin{equation}
\begin{array}{rlrlrl}
\hat{O} & &\stackrel{\cal W}{\longleftrightarrow} & ~& \tilde{o}(z) 
\nonumber \\
&\searrow \nwarrow & &~~ W\swarrow \nearrow & \nonumber \\
& & ~~~~~~~o(y,x) & & 
\label{trio.1}
\end{array}
\end{equation}
Note that $W$ is a well defined map by itself between a bilocal function 
$o(y,x)$ and the function $\tilde{o}(z)$. 
It can exist as a transformation independently from ${\cal W}$, a fact  
which we exploit later in sections D.3 and D.5. 

The noncommutativity of the coordinates in ${\cal Z}_{\star}$  
is encoded in the associative $\star$-product which is  
implicit in (\ref{action.5}). The $\star$-product is best described by 
the Weyl map 

$W: [\int\,du\,o_1(y,u)\,o_2(u,x)]~\leftrightarrow ~ 
 \tilde{o}_1(z) \star_{z}  \tilde{o}_2(z)$ where   
\begin{eqnarray}
\tilde{o}_1 \star_{z} \tilde{o}_2 &=& \tilde{o}_1
 \exp\{{i \over 2}
\stackrel{\gets}{\partial_{z_i}}\,\theta_{i\,j}
\stackrel{\to}{\partial_{z_j}})\} \tilde{o}_2 \nonumber \\
& \ne & \tilde{o}_2 \star_{z} \tilde{o}_1   
\label{action.5b}
\end{eqnarray}
with the arrows indicating the direction that the partial derivatives act.  
The measure of the noncommutativity of $\tilde{o}_1$ and $\tilde{o}_2$  
in ${\cal Z}_{\star}$ is provided by the Moyal bracket (MB)   
$\{\tilde{o}_1,\tilde{o}_2\}^{(M)}_z=
\tilde{o}_1 \star_z \tilde{o}_2-\tilde{o}_2 \star_z\tilde{o}_1$. 
We now consider for $\hat{O}$ the waveoperator $\hat{\rho}$. Then,  
$o(y,x)$ and $\tilde{o}(z)$ correspond to $\rho(y,x)$ and   
$\tilde{\rho}(z)$ respectively.  	
Using (\ref{action.5}) the Weyl map of (\ref{action.3}) is found to be  
\begin{equation}
i\hbar\,\partial_t\,\tilde{\rho}(z)=
\tilde{h}(z) \star_{z} \tilde{\rho}(z)~. 
\label{action.6}
\end{equation}
Here $\tilde{h}$ is the Weyl map of the bilocal Hamiltonian $h$ in 
 (\ref{action.3}). We refer to Eq.\,(\ref{action.6}) as 
the $\star$-Schr\"{o}dinger equation which is basically an operator relation
manifested by the waveoperator 
interpretation [Note that if $\hat{\rho}$ described the standard quantum 
mechanical density operator, 
the right hand side of (\ref{action.6}) would not be given by a single  
$\star$-product but by a MB instead. Hence (\ref{action.6}) would be  
the Moyal equation for the density operator 
 $i\hbar\,\partial_t\,\hat{\rho}=\tilde{h}\star \tilde{\rho} - 
\tilde{\rho}\star \tilde{h}=
\{\tilde{h},\tilde{\rho}\}^{(M)}$]. 

If $\tilde{\rho}$ is an existing solution of (\ref{action.6}), 
a new solution can 
be obtained by $\star$-multiplying that equation from the right by 
{\it time independent} function $\tilde{\xi}(z)$ such that 
$-i\hbar\,\partial_t\,\tilde{\xi}(z)=\tilde{\xi}(z) \star \tilde{h}=0$. 
The new solution is then represented by 
$\tilde{\rho}^\prime(z)=\tilde{\rho} \star_{z} \tilde{\xi}$ and respects 
$i\hbar\,\partial_t\,\tilde{\rho}^\prime=
\{\tilde{h},\tilde{\rho}^\prime\}^{(M)}$ with unique solutions (upto initial 
conditions). 

In the separable case, Eq.\,(\ref{action.6}) 
has also a specific physical realization in terms of the generalized Wigner 
function. Consider the  
case $\hat{\rho}=\vert \phi_{E}\rangle\,\langle 
\chi \vert$ where $E$ labels some basis states $\vert \phi_{E}\rangle$ 
which are the eigen solutions of the Schr\"{o}dinger equation with some 
Hamiltonian and $\vert \chi\rangle$ is a time independent  
state under the same Hamiltonian. In this separable form 
$\rho(y,x)=\phi_{E}(y)\,\chi^*(x)$ is an eigen solution of (\ref{action.3}) 
and the $\tilde{\rho}(z)$ is the corresponding eigen solution of 
(\ref{action.6}) which, by the application of (\ref{action.5}), is given by   
\begin{equation}
\tilde{\rho}(z)=\int\,dx\,e^{-i\,z_1\,x/\hbar}\,
\phi_{E}(z_2-{x \over 2})\,\chi^*(z_2+{x \over 2})~. 
\label{gwf}
\end{equation}
which is the generalized Wigner function\cite{CZ} ${\sl W}_{\chi,\phi_E}$.   
If the basis states $\vert \phi_{E}\rangle$ are the energy eigen basis of 
the Hamiltonian $\hat{\cal H}$ with energy $E$ then (\ref{gwf}) is the 
eigen solution of Eq.\,(\ref{action.6}).\cite{stargenvalue} 


Here $\rho(y,x)=\phi_{E}(y)\,\chi^*(x)$ is manifestly separable by choice, 
where the local solution is indicated by $\phi_{E}(y)$.
In order to explore beyond the ordinary local sector we  
must find how to reach the nonseparable sector of the doubled Hilbert 
space. 

\subsection{Separability and the Hilbert space operators}
We look for connections between the local $H \times H$ subsector 
and the general doubled Hilbert space.    
Hence the relevant question is changing the separability of the functions 
like $\rho(y,x)$. Within the frame of standard transformations in 
quantum mechanics
the answer is simple. Consider the Hilbert space operator 
-unitary or nonunitary- (speaking 
of the Hilbert space the standard -local- quantum mechanical Hilbert space is 
implied and not the doubled Hilbert space) 
$\hat{\Omega}$ acting on $\hat{\rho}$ as
\begin{equation}
\hat{\Omega}: \hat{\rho}=\hat{\rho}^\prime=\hat{\Omega}\,
\hat{\rho}\,\hat{\Omega}^{\dagger}
\label{noseparation.1}
\end{equation}
Suppose that we start from 
a separable case in (\ref{noseparation.1}) as 
$\hat{\rho}=\vert \psi_1\rangle\,
\langle \psi_2\vert$, e.g. $\rho(y,x)=\psi_1(y)\,\psi_2(x)$. Then 
$\hat{\rho}^\prime=
\hat{\Omega}\,\vert \psi_1\rangle\,\langle \psi_2\vert\,\hat{\Omega}^{\dagger}=
\vert \psi_1^\prime\rangle \,\langle \psi_2^\prime\vert$. 
where $\Vert \psi_j\Vert=\Vert \psi_j^\prime \Vert$. 
We observe that 
$\hat{\rho}^\prime$ is still separable, e.g. 
$\rho^\prime(y,x)=\psi_1^\prime(y)\,
\psi_2^\prime(x)$. Likewise, the 
invertibility of $\hat{\Omega}$ ensures that, the nonseparable 
$\rho(y,x)$ is transformed into a nonseparable $\rho^\prime(y,x)$.  
Therefore  
Eq.\,(\ref{noseparation.1}) is a simple proof that the separable   
and the nonseparable BLC representations are disjoint within the reach of 
standard transformation $\hat{\Omega}$. 
Our discussion here should give us the clue 
that in reaching for the nonseparable solutions one needs to go beyond  
$H \times H$. In searching for the answer, an 
extended view of the unitary (canonical) transformations in ${\cal Z}_\star$ 
will be given in the following section. 

\subsection{An extended view of quantum canonical maps}
One alternative
way to look into the unitary transformations is by canonical maps in
${\cal Z}_{\star}$.
Consider $\hat{z}_j$ and $\hat{Z}_j$ as the old and the new noncommutative 
generalized coordinates. A canonical map, in the operator picture, is given by 
$\hat{\Omega}: 
\hat{z}_j ~ \mapsto ~ \hat{Z}_j=Z_j(\hat{z}_1,\hat{z}_2)~,~~(j=1,2)$ 
such that the canonical commutation relations are preserved
\begin{equation}
\hat{\Omega}: [\hat{z}_j,\hat{z}_k]=[\hat{Z}_j,\hat{Z}_k]=i\,\theta_{j\,k}~.
\label{action.1b}
\end{equation}
The corresponding relations in ${\cal Z}_{\star}$ can be obtained by the 
Weyl transform and are based  
on the {\it canonical} MB, where the latter is defined by 
$\{z_i,z_j\}^{(M)}=i\,\theta_{ij}=\{Z_i,Z_j\}^{(M)}$. The {\it canonical} MB 
is {\it basis independent} [see Eq.\,(\ref{classcov.3}) below] 
whereas a general MB 
is not. Because of this, we will choose the $z$ basis in expressing 
the $\star$ products and MBs, 
hence $\star=\star_z$ as given by (\ref{action.5b}). 
Here  
$Z_j=Z_j(z_1,z_2)~,~~j=1,2$ are the Weyl symbols of the 
new noncommutative generalized coordinates $\hat{Z}_j$. In the following two 
different types of canonical maps will be examined  
as classical and quantum. These two types are distinguished 
by their covariance properties when they act on the functions of $z$.  
We refer to Ref.\,[16] for a detailed comparative study of the classical  
and the quantum cases.   

\subsubsection{Classically covariant canonical maps
$\Omega_{\cal Z}$: type-I} 
Since we are dealing with the 
classical case, a necessity now arises to differentiate 
the classical commutative space ${\cal Z}$ from the noncommutative one 
${\cal Z}_{\star}$. The commutativity in 
${\cal Z}$ is induced by the product  
$\tilde{o}_1(z)\, \tilde{o}_2(z)=\tilde{o}_2(z)\, \tilde{o}_1(z)$ 	
as opposed to the noncommutative one in ${\cal Z}_{\star}$ as expressed in 
(\ref{action.5b}). However there are also common features. 
For instance, a square integrable function in 
${\cal Z}_{\star}$ is also square integrable in ${\cal Z}$ due to the 
integral property of the $\star$-product $\int\,d^2z\,\tilde{o}_1(z) 
\star_z \tilde{o}_2(z)=
\int\,d^2z\,\tilde{o}_1(z) \,\tilde{o}_2(z)$. 
This allows us to refer to the 
functions of $z$ without referring to the underlying commutative or 
noncommutative space. The representations of the operators, however, 
generally differ due to the different 
covariance properties of them in ${\cal Z}$ or ${\cal Z}_{\star}$.  
 
With this in mind, we assume that an infinitesimal generator 
${\cal G}(z)$ of the canonical map exists generating the first order 
infinitesimal changes in the function 
$\tilde{o}(z)$, i.e. $\tilde{o} ~\to ~\tilde{o}+\delta\,\tilde{o}$, 
by the Poisson Bracket (PB). This is given by the classical 
textbook formula,  
$\delta\,\tilde{o}(z)=\epsilon\,\{{\cal G},\tilde{o}\}^{(P)}$ where a 
general PB 
is given between two such functions $\tilde{o}_1$ and $\tilde{o}_2$ by 
$\{\tilde{o}_1,\tilde{o}_2\}^{(P)}\equiv 
J_{j\,k}\,(\partial_{z_j}\,\tilde{o}_1)\,(\partial_{z_k}\,\tilde{o}_2)$.    
According to the standard phase space analytical mechanics, ${\cal G}$ is 
transformed into a Hamiltonian vector field as $X_{\cal G}=
J_{j\,k}\,(\partial_{z_j}\,{\cal G})\,(\partial_{z_k})$  
which generates the same infinitesimal change by the Lie bracket 
$\delta\,\tilde{o}(z)=\epsilon\,[X_{\cal G},\tilde{o}]=
\epsilon\,(X_{\cal G}\,\tilde{o}-\tilde{o}\,X_{\cal G})$. 
The classical canonical 
maps $\Omega_{\cal Z}$ we consider here are finite transformations  
in ${\cal Z}$ 
obtained by exponentiating the classical Hamiltonian vector fields as 
\begin{equation}
\Omega_{\cal Z}=e^{\epsilon\,X_{\cal G}}
\label{classgen.1}
\end{equation}
The finite action of $\Omega_{\cal Z}$ on the functions of $z$ is given by  
\begin{eqnarray}
\Omega_{\cal Z}: \tilde{o} &=& \tilde{o}+\epsilon\,[X_{\cal G},\tilde{o}]+
 \dots \nonumber \\ 
&+&  
{\epsilon^n \over n!}\,
\underbrace{[X_{\cal G},\dots[X_{\cal G}}_{n},[\tilde{o},X_{\cal G}]\dots] 
+\dots 
\label{classgen.2}
\end{eqnarray} 
which can be shown to have the manifest property
\begin{equation}
\Omega_{\cal Z}: \tilde{o}(z)=\tilde{o}(Z)
\label{classgen.3}
\end{equation}
where $Z=\Omega_{\cal Z}\,:\,z$. Another way of writing (\ref{classgen.3}) 
is 
\begin{eqnarray}
[\Omega_{\cal Z}: \tilde{o}_1(z)\,\tilde{o}_2(z)] &=& 
\tilde{o}_1^\prime(z)\,\tilde{o}_2^\prime(z) 
\nonumber \\
&=& \tilde{o}_1(Z) \, \tilde{o}_2(Z)~.
\label{classcov.1}
\end{eqnarray}
which will be a useful relation to facilitate the comparison with the 
quantum case. Eq's\,(\ref{classgen.1}-\ref{classcov.1}) are well-known  
results that can be found in the books on Lie algebraic techniques in 
analytical mechanics.\cite{Santilli}  
In (\ref{classcov.1}) 
the square brackets indicate that there are no other operators 
to the left or right acted upon by the expressions inside and 
the primes denote the transformed functions of $z$.  
This notation will be used throughout the paper. 
We note that Eq.\,(\ref{classcov.1})  defines the {\it classical covariance} 
as used in the context of this work.\cite{TH2} The action of the 
canonical map $\Omega_{\cal Z}$ on the Poisson Bracket (PB) can also be defined 
similarly. Denoting the latter by  
$\{\tilde{o}_1,\tilde{o}_2\}^{(P)} \equiv  
J_{j\,k}\,(\partial_{z_j}\,\tilde{o}_1)\,(\partial_{z_k}\,\tilde{o}_2)$, 
the classical covariance implies 
\begin{eqnarray}
[\Omega_{\cal Z}:  
\{\tilde{o}_1(z)&,&\tilde{o}_2(z)\}^{(P)}] \nonumber \\ 
&=& \{[\Omega_{\cal Z}:\tilde{o}_1],
[\Omega_{\cal Z}:\tilde{o}_2]\}^{(P)} 
\nonumber \\
&=&\{\tilde{o}_1(Z),\tilde{o}_2(Z)\}^{(P)}~. 
\label{classcov.1b}
\end{eqnarray}

\subsubsection{$\star$-covariant canonical   
maps $\Omega_{{\cal Z}_\star}$: type-II} 
Consider a unitary 
operator $\hat{\Omega}=e^{i\epsilon\,\hat{G}}$ with $\epsilon \in \mbox{\ee R}$ 
and $\hat{G}$ Hermitian. $\hat{\Omega}$ acts on the 
 operator $\hat{O}$ as
\begin{eqnarray}
\hat{\Omega}:\hat{O} &=& \hat{\Omega}\,\hat{O}\,\hat{\Omega}^\dagger 
\nonumber \\
&=& \hat{O}+i\,\epsilon\,[\hat{G},\hat{O}]+\dots 
\nonumber \\  
&+& {(i\epsilon)^n \over n!}\,
\underbrace{[\hat{G}\dots[\hat{G}}_{n},[\hat{O},\hat{G}]\dots]
+\dots 
\label{qgen.1}
\end{eqnarray}
In analogy with the classical case, $\hat{G}$ above is the generator of 
the quantum canonical maps in the Hilbert space. 
We describe the Weyl symbol of $\hat{\Omega}$ by 
$\omega_{{\cal Z}_\star}(z)$ and its adjoint $\hat{\Omega}^{\dagger}$ by 
$\omega_{{\cal Z}_\star}^{*}(z)$. 
Such a map is given by the commuting 
diagram
\begin{equation}
\begin{array}{rlrlrlrlrlrl}
& \hat{z}_j ~~ & \stackrel{\hat{\Omega}}{\longleftrightarrow}~~~
                                    &\hat{Z}_j &=&\hat{\Omega}\,\hat{z}_j\,
                                                       \hat{\Omega}^{\dagger}& 
\\
&{\cal W} \updownarrow &  ~~~~~~~  & \updownarrow {\cal W}& & ~  & \\
& z_j & \stackrel{\omega_{{\cal Z}_\star}}{\longleftrightarrow}  
~~~ & Z_j & =&\omega_{{\cal Z}_{\star}} : \, z_j &  
\end{array}
\label{diagram.1}
\end{equation}
It was shown\cite{TH2,TH1} that 
$\omega_{{\cal Z}_\star}$ acts on functions of $z$ as 
\begin{equation}
\omega_{{\cal Z}_\star}: 
\tilde{o}(z)=\tilde{o}^\prime(z)=
\omega_{{\cal Z}_\star}(z) \star_z \tilde{o}(z) 
\star_z \omega_{{\cal Z}_\star}^{*}(z)~.   
\label{starcov.1}
\end{equation}
The canonical map $\omega_{\cal Z_{\star}}$ is the quantum analog of the   
classical map of $\Omega_{\cal Z}$. We actually gained an advantage in the 
search for 
the methods to reach beyond the Hilbert space operator approaches by 
formulating the canonical maps in ${\cal Z}_{\star}$. The reason is that  
the quantum canonical maps can, in general, be handled 
in the phase space ${\cal Z}_{\star}$ in a similar way to the classical case 
in ${\cal Z}$ and this can be done in a totally independent way from the 
Hilbert space operator methods. Namely, the operator connections induced by 
the Weyl correspondence ${\cal W}$ can be totally and consistently ignored in 
(\ref{diagram.1}).Examining (\ref{trio.1}), what remains is the Weyl map $W$ 
between double (bilocal) Hilbert space and the phase space. We now    
proceed to discuss the $\star$-covariance in Eq.\,(\ref{starcov.1}). 

If we use two such functions 
$\tilde{o}_1$ and $\tilde{o}_2$, Eq.\,(\ref{starcov.1}) implies that 
\begin{equation} 
[ \omega_{{\cal Z}_\star}:  
 \tilde{o}_1 \star_z \tilde{o}_2] 
= [\omega_{{\cal Z}_\star}:\tilde{o}_1] 
\star_z [\omega_{{\cal Z}_\star}:\tilde{o}_2]~.
\label{starcov.2}
\end{equation}
Eq.\,(\ref{starcov.2}) is to be regarded as the extended version of the 
classical covariance in (\ref{classcov.1}). In the context of this work 
it will be referred to as the $\star$-covariance\cite{TH2}. 
Also note that, if we denote the canonical map in 
${\cal Z}_{\star}$ by 
$\omega_{{\cal Z}_\star}: z_j ~\to ~ Z_j$, Eq.\,(\ref{starcov.2}) 
implies violation of the classical covariance, e.g.   
$[\omega_{{\cal Z}_\star}:\tilde{o}](z) \ne \tilde{o}(Z)$. Comparing 
(\ref{classcov.1}) and (\ref{starcov.2}) we note that $\Omega_{\cal Z}$
in (\ref{classcov.1}) preserves the commutativity of the standard product 
between the functions $\tilde{o}_1$ and $\tilde{o}_2$ whereas for   
$\Omega_{{\cal Z}_\star}$ in (\ref{starcov.2}) the noncommutativity 
by $\star$-product
is invariant. The MB is therefore transformed by 
$\omega_{{\cal Z}_\star}$ as 
\begin{eqnarray}
[\omega_{{\cal Z}_\star}:  
\{\tilde{o}_1(z)&,&\tilde{o}_2(z)\}^{(M)} ] \nonumber \\ 
&=& \{[\omega_{{\cal Z}_\star}:\tilde{o}_1],
[\omega_{{\cal Z}_\star}:\tilde{o}_2]\}^{(M)} \nonumber \\
&\ne & \{\tilde{o}_1(Z),\tilde{o}_2(Z)\}^{(M)}~. 
\label{starcov.2b}
\end{eqnarray}
At the limit $\hbar~\to~ 0$ the classical and star covariances coincide.

\subsubsection{$\hbar$-independent canonical maps:}
A particular subgroup of canonical maps in types I and II is distinguished by 
its simultaneous representations both in ${\cal Z}$ and ${\cal Z}_{\star}$. 
This is the subgroup of $\hbar$-independent canonical maps and 
may be crucial for a unified understanding of the 
classical and quantum phase spaces. 

In what follows, we pay specific attention to this subgroup and 
derive some of its properties. Consider a canonically conjugated pair  
$Z_j=Z_j(z_1,z_2)~;~~(j=1,2)$ with 
$\{z_j,z_k\}^{(M)}=\{Z_j,Z_k\}^{(M)}=i\theta_{j\,k}$
where both the old $z_k$ and the new $Z_k$ canonical variables are assumed
to be independent of $\hbar$. 
Expanding the MB in $\theta_{j\,k}=\hbar\,J_{j\,k}$ we have 
\begin{eqnarray}
\{Z_j,Z_k\}^{(M)}_{z}&=&i\hbar\,\{Z_j,Z_k\}^{(P)}_{z}+{\cal O}(\theta^3)  
\label{classcov.2}
\end{eqnarray}
where the first term is the Poisson Bracket (PB) 
$\{Z_j,Z_k\}^{(P)}_{z} \equiv J_{\ell\,m}\,
(\partial_{z_\ell}Z_j)(\partial_{z_m}Z_k)$.   
Since we assumed that $Z_j$'s have no dependence on $\hbar$, 
by equating (\ref{classcov.2}) to $i\theta_{j\,k}$ and matching the powers 
of $\theta$ we deduce that the 
${\cal O}(\theta^3)$ and higher order terms in 
(\ref{classcov.2}) must all vanish. The nonvanishing part of 
(\ref{classcov.2}) is   
\begin{eqnarray}
\{Z_j,Z_k\}^{(M)}_{z} &=& i \hbar \, \{Z_j,Z_k\}^{(P)}_{z} \nonumber \\   
&=& i\hbar J_{j\,k}
\label{classcov.3}
\end{eqnarray}
which means that $\{Z_j,Z_k\}^{(P)}_{z}=J_{j\,k}$. This is the condition for 
the pair $Z_1,Z_2$ to be classically canonical. 
We learn that any $\hbar$ independent quantum canonical map   
$\omega_{{\cal Z}_\star}: z_j ~ \to ~ Z_j$   
in (\ref{classcov.2}) implies (\ref{classcov.3}). Therefore such maps 
are also canonical in the Poisson sense and conversely all $\hbar$-independent 
Poisson maps are also quantum canonical. We will refer to them as type-I  
as well. The corollary of this result is that  
an $\hbar$ independent quantum canonical pair can also be obtained 
by an appropriate $\Omega_{\cal Z}$. 
Now, we consider a second $\hbar$-independent 
map $\Omega_{\cal Z}^\prime$. By definition it acts classically 
on the canonical MB as   
\begin{eqnarray}
\Omega_{\cal Z}^\prime: {1 \over i\hbar}\{Z_j,Z_k\}^{(M)}&=& 
\Omega_{\cal Z}^\prime: \{Z_j,Z_k\}^{(P)} \nonumber \\
&=& \{[\Omega_{\cal Z}^\prime: Z_j],
[\Omega_{\cal Z}^\prime: Z_k]\}^{(P)} \nonumber \\
&=& \{Z_j(z^\prime),Z_k(z^\prime)\}^{(P)} \nonumber \\
&=& J_{j\,k}
\label{classcov.4}
\end{eqnarray}
where $z^\prime=\Omega_{\cal Z}^\prime: z$. 
In (\ref{classcov.4}) the second line 
is obtained by the classical covariance of the Poisson bracket in 
(\ref{classcov.1b}). The third line is an application of (\ref{classcov.1}). 
The last line is the statement of the invariance of the canonical MB under 
$\hbar$-independent canonical maps which is a corollary of (\ref{classcov.3}). 
Therefore the classically covariant [see (\ref{classcov.1})] canonical maps 
define an $\hbar$-independent automorphism on the 
{\it canonical} MB. That this result is correct only for canonical
pairs and not for arbitrary functions of $z$ is the essence of the
$\star$-covariance. In the literature, it almost goes  
without saying that all quantum canonical maps fall in the type-II. As we 
have seen above, type-I maps also preserve the canonical MB although their 
action is truly different from that of type-II. Therefore the space of 
canonical maps in ${\cal Z}_\star$ is actually a union of types I and II  
which is referred to as the extended quantum phase space.  

\subsubsection{Nonseparability and the type-I maps}
In the standard quantum mechanics, the canonical transformations of type-II 
have integral transforms. Denoting by $\phi(x)$ some local Hilbert space  
function, a typical map $\hat{\Omega}$ on $\phi(x)$ is expressed by 
\begin{equation}
(\hat{\Omega}: 
\phi)(x^\prime)=\int\,dx^\prime\,u(x,x^\prime)\,\phi(x^\prime)~.  
\label{intrep}
\end{equation} 
In the rest of this letter, we will be interested in examining  
similar integral transforms   
 adopted for the bilocal representations and for canonical  
maps of type-I. For separable cases these bilocal transforms reduce to 
the direct products of the integral transforms like in 
(\ref{intrep}) of which an example is given below from the linear 
canonical group. Denoting by $\Omega_{\cal Z}$ a generic 
type-I map, its action can be written as  
\begin{eqnarray} 
o^\prime(y,x)&=&(\Omega_{\cal Z}:o)(y,x) \nonumber \\
&=&\int {d^2 z \over 2\pi\hbar}\,{\cal K}_z(y,x)\,
\tilde{o}^\prime(z)
\label{separate.1}
\end{eqnarray}
with $\tilde{o}^\prime(z)=(\Omega_{\cal Z}:\tilde{o})(z)=
\tilde{o}(Z)$ as dictated by (\ref{classcov.1}). 
Using the inverse of (\ref{action.5})
we convert (\ref{separate.1}) into an integral transform between
the old and the new functions of BLC as 
\begin{equation}
o^\prime(y,x)=\int du\,\int\,dv\,
{\cal L}_{\Omega_{\cal Z}}(y,x;v,u)\,o(v,u)
\label{separate.2}
\end{equation}
with the integral kernel 
\begin{equation}
{\cal L}_{\Omega_{\cal Z}}(y,x;u,v)=\int\,{d^2 z \over 2\pi\hbar}
\,{\cal K}_z(y,x)\,
(\Omega_{\cal Z}: {\cal K}_z)(v,u)~.  
\label{separate.3}
\end{equation}
In Eq.\,(\ref{separate.3}) 
the classical covariance in (\ref{classcov.1}) implies that  
$\Omega_{\cal Z}:{\cal K}_z={\cal K}_{\Omega_{\cal Z}: z}={\cal K}_{Z}$. 
Eq.\,(\ref{separate.2}) is the bilocal extension of (\ref{intrep}).  
We now examine the separability of the general canonical map in
(\ref{separate.2}) by three general examples.

a) Linear canonical group  
$\Omega_{\cal Z}^{(a)}=\Lambda_g \in Sp_2(\mbox{\ee R})$ acts on the 
phase space as 
\begin{equation}
\Lambda_g: \tilde{o}(z)=\tilde{o}(g^{-1}:z) ~,\qquad 
g=\pmatrix{a & b\cr c &d\cr}
\label{sp2r.1}
\end{equation}
where $\det g=1$ and $g^{-1}:z=Z$ is the transformed coordinate. 
It is known that the linear canonical group is the only group of 
transformations for which the  
classical (type-I) and $\star$-covariances (type-II) coincide. Based on
this fact we already expect (\ref{sp2r.1}) not to have an effect
on the separability. Nevertheless, the explicit calculation is illustrative. 
Using the inverse of the Weyl kernel
in (\ref{action.5}) we map the transformed and initial solutions by
calculating the kernel in (\ref{separate.3}) for $\Lambda_g$ as 
\begin{eqnarray}
{\cal L}_{g}&=&
\int\,{d^2 z \over 2\pi\hbar}\,
{\cal K}_z(y,x)\,(\Lambda_g: {\cal K}_z)(v,u) \nonumber \\
&=&
\int\,{d^2 z \over 2\pi\hbar}\,{\cal K}_z(y,x)\,{\cal K}_{g^{-1}:z}(v,u)  \\
&=& \exp\{{i \over 2\hbar\,c}[d(u^2-v^2)+
a(x^2-y^2)-2(xu-yv)]\} \nonumber \\ 
\label{sp2r.2}
\end{eqnarray}
Here ${\cal L}_g$ is the kernel of the map $\Lambda_g$. As expected it    
is manifestly separable, i.e. ${\cal L}_g(y,x;u,v)={\cal L}_g(0,x;u,0)\,
{\cal L}_g(y,0;0,v)$. Therefore the transformed solution is separable when  
$o(y,x)=\psi_1(y)\,\psi_2(x)$ and nonseparable when $o(y,x)$ is nonseparable.  
In the first case both dependences on the
coordinates are transformed separately and identically as 
\begin{eqnarray} 	
Y^\prime(y)\,X^\prime(x)&=&
\Bigl[\int\,dv\,{\cal L}_g(y;v)\,Y(y)\Bigr]\,
\nonumber \\
&\times &\Bigl[\int\,dv\,{\cal L}_g(x;u)\,X(x)\Bigr]~.
\label{sp2r.2b}
\end{eqnarray}
where each square bracket is an integral transform of type-II as in 
(\ref{intrep}). 

b)  As the second kind of type-I maps we consider    
\begin{eqnarray}
\Omega_{\cal Z}^{(b)}:&& \tilde{o}(z) 
= 
\tilde{o}(z_1+\tau_1 \,A_1(z_2),z_2+\tau_2\,A_2(z_1))  
\label{gauge.1}
\end{eqnarray}
where it is crucial that the functions $A_j$'s are $\hbar$
independent. The free and real parameter $\tau_j$ is either momentum,  
($\tau_1=\tau, \tau_2=0$) or coordinate-like  
($\tau_1=0, \tau_2=\tau$).
Two or higher dimensional versions of (\ref{gauge.1})
are normally referred to as the gauge transformations (not considered  
here). Here we consider a  momentum-like map ($\tau_1=\tau \ne 0,
\tau_2=0$). Using (\ref{gauge.1}) in (\ref{action.5}) one finds    
\begin{equation}
o^\prime(y,x)=e^{i{\tau \over \hbar}\,x_{-}\,A(x_+)}\,o(y,x) 
\label{gauge.2}
\end{equation}
where $x_-=(x-y)$ and $x_+=(x+y)/2$. 
Due to the exponential factor 
the transformation kernel in (\ref{gauge.2}) is, for    
general cases, manifestly nonseparable. Exception is when 
$A(x_+) \propto \alpha+\beta \,x_+$, with $\alpha,\beta$ constant,  
in which case the exponential separates. 

c) As the last general example of canonical maps we examine the  
classical change of variables  
\begin{equation}
\Omega_{\cal Z}^{(c)}: \tilde{o}(z) ~= ~ 
\tilde{o}(T(z_1),z_2/T^\prime(z_1))
\label{cov.1}
\end{equation}
or $1 ~\leftrightarrow ~2$ on both sides. Here $T(z_1)$ describes an 
invertible and $\hbar$-independent function. 
Using (\ref{cov.1}) in (\ref{action.5}) one finds  
\begin{eqnarray}
o^\prime(y,x) &&= T^\prime(x_+) \nonumber  \\
&\times&~o(T(x_+)+{x_- \over 2}T^\prime(x_+) 
,T(x_+)-{x_- \over 2}T^\prime(x_+)) \nonumber \\ 
\label{cov.2}
\end{eqnarray}
This example is also manifestly nonseparable for general cases. 

These two type of maps in (b) and (c) can transform a local 
(separable) picture into a nonlocal (nonseparable) one. 

Any nonseparable canonical map can have various compositions 
of elementary maps of type-II but it must also  
must include type-I maps such as $\Omega^{(b)}_{\cal Z}$ and/or 
$\Omega^{(c)}_{\cal Z}$. 

Concerning the type-II case, these are standard unitary transformations of  
which an example we discussed as $Sp_2(\mbox{\ee R})$ [part (a) above]. 
The type-II maps within the standard Weyl formalism 
can be represented in the Hilbert space operator form as in 
(\ref{noseparation.1}) and the results obtained 
therein are valid for them. For the purpose of separability 
these standard maps do not offer interesting results by themselves. 

\subsubsection{Unitarity of the type-I maps}
Coming back to type-I maps, our discussion leading to Eq.\,(\ref{classcov.3}) 
classifies them also as quantum canonical maps. An interesting question is then 
how to incorporate them into the quantum mechanical picture. To the author's 
knowledge, the type-I maps have been first examined    
in a quantum mechanical context by the authors of Ref.\,[19]. Instead of  
the phase space, these authors searched for the standard Hilbert space 
representations. The equivalence genereted by the type-I maps is known as  
isometry, viz. they map a Hilbert space to an equivalent Hilbert space with 
a normally different inner product. This point has been also advocated more 
recently by Anderson\cite{Anderson}. 

In the phase space, due to their classical origin the type-I maps act on 
${\cal Z}$. They can nevertheless be incorporated as unitary  
transformations in ${\cal Z}_{\star}$. Therefore the function space 
${\cal L}_2({\cal Z})$ becomes the appropriate 
Hilbert space for such maps. That they conserve the norm of functions in 
${\cal L}_2({\cal Z})$, 
and hence the unitarity, can be shown by using (\ref{action.5}) and 
its inverse as,      
\begin{eqnarray}
\int\,dy\,dx\,\vert o^\prime(y,x)\vert^2  
& \equiv & \int\,{d^2 z \over 2\pi\hbar}\,
\tilde{o}^{\prime^*}(z) \star_z 
\tilde{o}^\prime(z) 
\nonumber \\
&=& \int\,{d^2 z \over 2\pi\hbar}\,
\tilde{o}^*(\Omega_{\cal Z}:z) \star_z \tilde{o}(\Omega_{\cal Z}:z)
\nonumber \\
&=& \int\,{d^2 z \over 2\pi\hbar}\,
\tilde{o}^{*}(Z) \star_z \tilde{o}(Z)
\nonumber \\
&=&\int\,{d^2 z \over 2\pi\hbar}\,
\tilde{o}^{*}(Z) ~ \tilde{o}(Z)
\nonumber \\
&=&\int\,{d^2 Z \over 2\pi\hbar}\,
\tilde{o}^{*}(Z) ~ \tilde{o}(Z)
\nonumber \\
&=&\int\,{d^2 Z \over 2\pi\hbar}\,
\tilde{o}^{*}(Z) \star_Z \tilde{o}(Z) \nonumber \\
&\equiv &\int\,dy\,dx\,\vert o(y,x)\vert^2~.  
\label{norm.1}
\end{eqnarray}
The preservation of the norm in ${\cal L}_2({\cal Z})$ follows by comparing 
the top and the second line from the bottom. In the first line     
we used $\tilde{o}^\prime(z)=(\Omega_{\cal Z}: \tilde{o})(z)$. 
The second and the third lines are the application of classical covariance in 
(\ref{classcov.1}). The fourth line is  
the general property of the $\star$-product, i.e. $\int\,d^2 z 
\tilde{o}_1 \star \tilde{o}_2=
\int\,d^2 z\,\tilde{o}_1 \tilde{o}_2$. 
In the fifth line the 
invariance of the integral measure, i.e. $d^2 z=d^2 Z$ is employed. The sixth 
line is the general property of the $\star$-product again].  

\subsection{Nonlocality, entanglement and generalized Bell states} 
The key reason why a general type-I transformation is not representable 
locally is that such a map connects a local representation  
to a nonlocal one and, in this respect it can be viewed as an isometry 
between two Hilbert spaces which we write formally as  
$\hat{\Omega}: H \to H^\prime$
, where the Hilbert spaces $H$ and $H^\prime$ are distinguished 
by their different inner products\cite{Anderson}. Consider two copies 
$H_x$ and $H_y$ of the  
same Hilbert space and an operator defined in $H_x \times H_y$. 
In view of section D.4, a type-I map acting on this operator     
entangles the local coordinates   
and the transformed operator does not have 
a direct product representation.  
The important point is that the same map  
can be represented unitarily in an appropriate two-dimensional  
noncommutative space. In the context of this work the 
noncommutative space in question is the phase space ${\cal Z}_{\star}$. 
How $\hat{\Omega}$ induces entanglement/disentanglement in terms  
of the bilocal coordinates can be further demonstrated in the following 
example. Consider for instance the operator     
$\vert n\rangle\,\langle m\vert$, where $\vert n\rangle$ denotes  
the harmonic oscillator energy eigenstates, represented bilocally   
in $H_x \times H_y$ or equivalently, 
as a generalized Wigner function\cite{CZ}  
$\tilde{\sl W}_{n,m}(z)$. A type-I  
map acting on this operator can be defined in the bilocal representation    
by using (\ref{separate.2}) and (\ref{separate.3}) as 
\begin{equation}
(\hat{\Omega}:\vert n\rangle\langle m\vert)(y,x)=\int\,du\,\int\,dv\,
{\cal L}(y,x;v,u)\,\psi^*_n(y)\,\psi_m(x)
\label{entangl.1}
\end{equation}
where $\psi_m(x)=\langle x\vert n\rangle$ is the m'th harmonic oscillator 
Hermite Gaussian and 
\begin{eqnarray}
{\cal L}(y,x;v,u)&=&\int{dz_1 \over 2\pi\hbar}\,
e^{-i[Z_1(z_1,x_+)u_- -z_1 x_-]}\,\nonumber \\
&\times& \delta(Z_2(z_1,x_+)-u_+)  
\label{entangl.2}
\end{eqnarray}
where $x_+=(x+y)/2,~x_-=(x-y),~u_+(u+v)/2,~u_-=(u-v)$. Equivalently, 
$\Omega_{\cal Z}$ acts on $\tilde{W}_{n,m}(z)$ in the phase space.   
Using the results in section D.1 we write this as 
$\Omega_{\cal Z}:\tilde{\sl W}_{n,m}(z)=\tilde{\sl W}_{n,m}(\Omega_{\cal Z}:z)
=\tilde{\sl W}_{n,m}(Z)=\tilde{\sl W}^\prime_{n,m}(z)$ where $Z$ standardly 
denotes the transformed phase space variables. The $\tilde{\sl W}_{n,m}(z)$'s  
form a basis (the Wigner basis) in the phase space 
via the orthogonality relation 
$\int\,d^2\,z/2\pi\hbar\,\tilde{\sl W}_{n,m}^*(z)\,
\tilde{\sl W}_{n^\prime,m^\prime}(z)=\delta_{n,n^\prime}\,
\delta_{m,m^\prime}$. The transformed Wigner function can be expanded  
in this basis as 
\begin{eqnarray}
\Omega_{\cal Z}: {\sl W}_{n,m}(z)&=& 
\sum_{n^\prime,m\prime}\,\omega_{n^\prime,m^\prime}^{(n,m)}\,
\tilde{\sl W}_{n^\prime,m^\prime}(z) \nonumber \\
\omega_{n^\prime,m^\prime}^{(n,m)}&=&\int\,{d^2 z \over 2\pi\hbar}\,
\tilde{\sl W}_{n,m}^*(z)\,\tilde{\sl W}_{n^\prime,m^\prime}(Z)~. 
\label{entangl.3}
\end{eqnarray}
Let us consider a simple case  
$\tilde{\sl W}_{n,m}=\delta_{n,m}\, \tilde{\sl W}_{n}$. From a physical 
point of view this represents a specific pure (number) state density matrix.  
The corresponding Wigner function is rotationally invariant in 
${\cal Z}_{\star}$ and is given by the well known expression\cite{HG} 
\begin{equation}
\tilde{\sl W}_{n}(z)=2\,(-1)^n\,
e^{-(z_1^2+z_2^2)}\,{\sl L}_n(2(z_1^2+z_2^2))~.  
\label{entangl.3b}
\end{equation} 
where ${\sl L}_n$ is the n'th Laguerre polynomial. 
The rotational invariance in (\ref{entangl.3b}) 
is implied by the zero eigenvalue of the 
{\it phase space angular momentum} operator  
$K_0=i\hbar\,z_j\,\theta_{j,k}\,\partial_{z_k}$ viz., 
$K_0\,: \tilde{\sl W}_n(z)=0$ [a more general fact is  
$K_0\,:\tilde{\sl W}_{n,m}=(n-m)\hbar\,\tilde{\sl W}_{n,m}$]. Let us further 
assume that $[\Omega_{\cal Z},K_0]=0$. Then the transformed  
Wigner function ${\sl W}_{n}(Z)$ 
is also rotationally invariant. It can therefore  
be expanded in a rotationally invariant Wigner subbasis 
\begin{equation}
\tilde{\sl W}_{n}(Z)=\sum_{n^\prime}\,\omega_{n^\prime}^{(n)}\,
\tilde{\sl W}_{n^\prime}(z) 
\label{entangl.4}
\end{equation}
where the normalization on both sides and the invariance of the 
measure $d^2z=d^2Z$ requires  
$\sum_{n^\prime}\,\omega_{n^\prime}^{(n)}=1$. The local symmetry algebra of 
(\ref{entangl.4}) is $sp_2(\mbox{\ee R})$. 
The nonlocal symmetries are generated by 
the remaining type-I generators forming an infinite Lie subalgebra of the   
canonical algebra\cite{Anderson,LS}. 
A general rotationally invariant transformation as in 
(\ref{entangl.4}) can be generated in two steps 
where the first step is 
$T_1: (z_1,z_2) ~\to ~(J,\theta)$ with $(J,\theta)$ describing the   
generalized action-angle coordinates where for this example they specifically 
are $J=(z_1^2+z_2^2)$, $\theta=\tan^{-1}z_1/z_2$. The second step is  
$T_2: (J,\theta) \to (g(J),\theta/g^\prime(J))$ where we require $g(J)$ to   
be of type-I hence it is $\hbar$ independent. The final transformation 
$T_2\,T_1$ produces (\ref{entangl.4}) and it can be written\cite{LS} 
as a composition of the generators in the canonical Lie subgroup and  
$SP_2(\mbox{\ee R})$. Once the contact map $g(J)$ is known, the 
coefficients $\omega^{(n)}_{n^\prime}$ are calculated by 
[from (\ref{entangl.3b}) and (\ref{entangl.4})]  
\begin{equation}
\omega_{n^\prime}^{(n)}=2\,(-1)^{n+n^\prime}\,\int_{0}^{\infty}
 dJ\,e^{-(J+g(J))}\,
{\sl L}_{n^\prime}(2J)\,{\sl L}_n(2g(J))~. 
\label{entangl.5}
\end{equation}
Eq.\,(\ref{entangl.5}) characterizes a general nonlocal, radial 
(rotationally invariant) transformation. These transformations generate   
the bilocal Hilbert space analogs of the standard (entangled) Bell states  
as shown below. Hence they can be referred to as 
 {\it generalized~Bell~states}. Here 
we continue with a specific example of a generalized Bell state. 
The simplest one has two nonzero coefficients in 
(\ref{entangl.4}) as      
$\omega^{(n)}_{n^\prime}=a_{n,n_1}\,\delta_{n^\prime,n_1}+ a_{n,n_2}\,
\delta_{n^\prime,n_2}$. Now let us find a canonical map that generates   
this Bell state. Consider the specific case $n=0$ with  
$n_1=0$ and $n_2=2$ in (\ref{entangl.5}). For $n=0$ one of 
the Laguerre polynomials in (\ref{entangl.5}) drops out (${\sl L}_0=1$). 
We next choose 
$g(x)=x-\ln[a_{0,0}+a_{0,2}\,{\sl L}_2(2x)]$ (here we consider 
$a_{0,2} < a_{0,0}$ so that the logarithm is real). Using this in 
(\ref{entangl.5}) it can be seen that 
\begin{equation}
\vert 0\rangle\,\langle 0\vert ~ \to ~
a_{0,0}\,\vert 0\rangle\,\langle 0\vert 
+a_{0,2}\,\vert 2\rangle\,\langle 2\vert 
\label{bell.1}
\end{equation}
is obtained. More general polynomial or infinite series Bell 
states can also be obtained. For instance, consider for $n=0$ the map  
$g(x)=x-\ln[\sum_{n^\prime=0}^{M_{max}}\omega^{(0)}_{n^\prime}\,
{\sl L}_{n^\prime}(2x)]$ where $M_{max}$ can be finite or infinite.  
This produces the map   
\begin{equation}
\vert 0\rangle\,\langle 0\vert ~ \to ~ 
\sum_{n^\prime}^{M_{max}}\,a_{0,n^\prime} 
\vert n^\prime\rangle\,\langle n^\prime\vert~.
\label{bell.2}
\end{equation}  
Even Laguerre polynomials are bounded from below. This insures that it is 
always possible to choose a well defined (single valued and real) $g(x)$ by 
choosing the constant term $a_{0,0}$ appropriately and requiring  
the leading Laguerre function in the polynomial $M_{max}$ to be of even 
order. 

\subsubsection{Generalized Bell states as noncommutative solitons} 
Recently radially symmetric nonlocal solutions similar to 
(\ref{entangl.4}) have been observed (see for instance the 
first reference in Ref.\,[23] for a survey) in 
the noncommutative field theories as vacuum soliton solution in the infinite 
noncommutativity limit. In the context of these works 
$\vert 0\rangle\,\langle 0\vert$ is 
an example of a level-one noncommutative soliton and Eq.\,(\ref{bell.2}) 
is an example of a unitary map between two vacuum configurations 
of a level-one 
soliton. The type-I and type-II transformations are joined in the 
canonical group of area preserving diffeomorphisms in the phase space.  
Therefore, we expected that this canonical group is related to the   
$U(\infty)$ symmetry\cite{gopa}. In the language of Ref.\,[23] the 
type-II maps    
are identified with the {\it local} whereas the type-I maps are associated  
with the {\it nonlocal} sectors in $U(\infty)$. Whether the canonical group 
covers this $U(\infty)$ entirely is currently under investigation. 
Furthermore, the type-I maps are fundamentally different from the 
non-unitary isometries presently discussed in the literature. For instance 
Harvey\cite{Harvey} studied the non-unitary {\it phase~operator}   
$\hat{S}=\sum_{n=0}^{\infty}\,\vert n\rangle\,\langle n+1\vert$ in the context 
of noncommutative field theories as a generating map of fixed-level index 
noncommutative soliton solutions.  
The fact that $\hat{S}$ is representable in a local Hilbert space implies   
that its action preserves the local (nonlocal) sector; viz. 
${\cal H}_x \times {\cal H}_y ~\to ~ {\cal H}_x^\prime \times 
{\cal H}_y^\prime$. From the 
arguments above it is clear that such maps cannot create entanglement 
therefore they cannot induce transformations  
as in (\ref{bell.2}). In this view one immediate use of type-I 
transformations in the noncommutative field theory is in the generation of 
entangled vacuum soliton configurations. When the type-I maps can be used in 
composition with the $\hat{S}$ operator further soliton configurations can   
be obtained. 
As a typical composition of the two consider for instance the action of 
$\hat{S}^{\dagger}$      
on (\ref{bell.2}) one finds a transformation from  
$\vert 1\rangle\,\langle 1\vert$ to the generalized Bell state 
$\sum_{n^\prime}^{M_{\max}}\,a_{0,n^\prime}
\vert n^\prime+1\rangle\,\langle n^\prime+1\vert$
where leading term $M_{max}+1$ has an odd Laguerre polynomial. 
  
Now we briefly consider the radially nonsymmetric configurations. 
We start by identifying 
$K_0=i\hbar\,z_j\,\theta_{j,k}\,\partial_{z_k}=h_0\star-\star h_0$ 
where $h_0=(z_1^2+z_2^2)/2$ 
is the harmonic oscillator Hamiltonian, as the third generator of  
$sp_2(\mbox{\ee R})$. The left  
and the right $\star$ multiplication is equivalent to a doubling of the 
degrees of freedom whence, $K_0$ has an infinitely degenerate eigenbasis 
$\vert n+k\rangle\langle n\vert$ with eigenvalue $k$ for all $0\le n$. The 
other two generators of $sp_2(\mbox{\ee R})$, namely  
$K_{\pm}$ raise and lower $k$ by unity for all $n$ and, 
they are used as generators of non-unitary symmetries. These type of 
generators break the rotational symmetry by introducing non-zero phase space 
{\it angular momentum} $k$ and, in the context of Ref.\,[23], they 
generate soliton solutions with fixed angular momentum. 
We will examine the relations between the nonlinear canonical group and  
the $U(\infty)$ in this context in a separate work\cite{THCD}.  

\subsection{Discussion}
The general approach to the canonical maps in general, and the type-I maps 
particularly, require reaching beyond the standard Hilbert space formalism. 
Type-I canonical maps have been so constructed can generate Darboux 
transformations between two partner Hamiltonians\cite{Anderson2}. In this 
context they play role in mapping one integrable system to a large set of 
its integrable partners. In addition to these results, the current work 
demonstrates that a noncommutative, nonlocal    
extension of the quantum mechanics can be obtained from the standard  
one by the use of type-I as well as type-II canonical maps in an extended 
phase space. Type-I maps establish unitary isomorphism in the phase space 
and isometry in the Hilbert space and, they characterize the nonlocal 
formulation of quantum mechanics.  
This result can be illustrated in the following figure.   
\begin{figure}
~~~~~~~~~\psfig{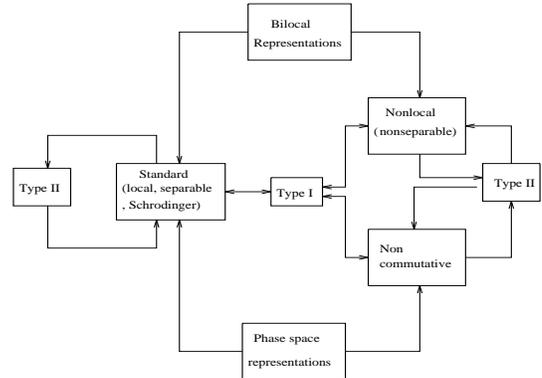}
\caption{Schematic of the bilocal and the phase space representations 
in quantum mechanics and their interconnections by 
the type I and II canonical maps.}
\end{figure}
In the view of this figure,
the standard representations are specific cases of, and can be obtained   
from, the bilocal ones 
at one end and the phase space representations at the other. The bilocal 
and the phase space representations are 
connected by a $W$ map which is not shown there. The second case 
that one can obtain from the bilocal ones is the nonlocal (nonseparable in 
BLC) representations. These are connected to the noncommutative picture of 
the phase space representations by the same Weyl map that connects the general 
bilocal and the phase space ones. Each (local and nonlocal) 
representation is an independent 
automorphism created by the type II maps. 
The type I maps join these otherwise disjoint representations.  

Note that, in the context of this work,
 the figure above resulted from a 
quantum mechanical analysis. One trivial 
extension is to carry out the analysis in 
$N$ coordinate ($2\,N$ phase space) dimensions. We have indications that 
for $1 < N$ the linear canonical group has nonlocal realizations\cite{THCD}. 
More interestingly,   
it has also implications for the field theories on noncommutative spaces.
In particular, the field equations in such theories are reminiscent of  
the $\star$-Schr\"{o}dinger equation in (\ref{action.6}) with the nonlinear 
field interactions added. The representations of these theories in the 
noncommutative space ${\cal Z}_{\star}$ as well as in the BLC can be fitted  
manifestly 
in the context of Fig.1. It is also specifically shown how nonlocal 
maps generate 
generalized Bell states. Interesting explorations of such maps exist in the 
generation and characterization of entangled soliton configurations 
in the noncommutative theories and in nonlocal quantum mechanics. 

\subsection*{Acknowledgments}
The author is thankful to C. Zachos and D. Fairlie and C. Deliduman  
for helpful discussions and critical comments. 


\begin{thebibliography}{99}
\bibitem{remark1} The waveoperator $\hat{\rho}$ here resembles in some sense   
the density operator in standard quantum mechanics but is not equal to it.  
Note that neither the crucial positivity condition nor the Hermiticity 
of the quantum mechanical density operator   
are imposed here and general results for $\hat{\rho}$ will be obtained 
without these conditions. However separable solutions  
$\rho=\psi_1\,\psi_2$ can be considered as the matrix elements of the 
density operator in some nondiagonal basis. These are connected to a bona-fide 
Wigner function by the Weyl map in (\ref{action.5}). We discuss this point 
later in Eq.\,(\ref{gwf}).
\end{thebibliography}
\end{document}